\documentclass[reprint,amsmath,amssymb,nofootinbib,aps,pra]{revtex4-2}
\usepackage{graphicx}
\usepackage{dcolumn}
\usepackage{bm}
\usepackage[breaklinks=true,colorlinks=true,linkcolor=blue,urlcolor=blue,citecolor=blue]{hyperref}
\usepackage{multirow}
\usepackage[dvipsnames]{xcolor}
\usepackage[pscoord]{eso-pic}
\usepackage[normalem]{ulem}
\usepackage{slashed}
\usepackage{makecell}

\newcommand{\placetextbox}[3]{
	\setbox0=\hbox{#3}
	\AddToShipoutPictureFG*{
		\put(\LenToUnit{#1\paperwidth},\LenToUnit{#2\paperheight}){\vtop{{\null}\makebox[0pt][c]{#3}}}
	}
}

\newcommand{\lya}{\text{Lyman-$\alpha$}~}

\newcommand{\tquote}[1]{``#1''}

\def\mnras{MNRAS}
\def\aj{AJ}

\def\prd{Phys. Rev. D}         % Physical Review D
\begin{document}

\placetextbox{0.90}{0.97}{\small ULB-TH/21-15}

\title{Hints of dark matter-neutrino interactions in Lyman-$\alpha$ data}

\author{
	Deanna C. Hooper,$^{1,2}$ and
	Matteo Lucca$^{1}$
	\\
	$^{1}$\textit{\small Service de Physique Th\'{e}orique, Universit\'{e} Libre de Bruxelles, C.P. 225, B-1050 Brussels, Belgium}
	$^{2}$\textit{\small Department of Physics and Helsinki Institute of Physics, PL 64, FI-00014 University of Helsinki, Finland}\\
}

\begin{abstract}
	In this work we investigate the possibility that dark matter and (massive) neutrinos can interact via a simple, constant cross section. Building on previous numerical efforts, we constrain this model with CMB, BAO and, in particular, \lya data.~We find that the latter hint to a significant departure from $\Lambda$CDM, with a preference for an interaction strength about 3$\sigma$ away from zero. We trace the origin of this preference back to the additional tilt that the interacting scenario can imprint on the \lya flux power spectrum, solving a well-known tension in the determination of this quantity between early-time and \lya probes. Future work including complementary \lya data as well as dedicated numerical simulations will be crucial in order to test these results.
\end{abstract}

\maketitle

\section{Introduction}
Two of the most persistent and prevalent mysteries of modern physics involve the fundamental nature of neutrinos and of dark matter (DM). The former, although phenomenologically understood in their role in the standard model (SM) of particle physics, still come with several open questions regarding, for instance, the origin of their mass and mass splitting (see \cite{Hernandez:2017txl} for a pedagogical review). DM, on the other hand, remains elusive in many searches (see e.g., \cite{Undagoitia:2015gya, Gaskins:2016cha, Kahlhoefer:2017dnp} for possible reviews of direct, indirect, or collider searches, respectively), and the available mass range covers over 70 orders of magnitude (see e.g., Fig. 1 of \cite{Battaglieri:2017aum}). The reason for the puzzling nature of these species relies on their (yet unexplained) extremely weak interaction strengths with other particles -- indeed it is not even known if the DM has any fundamental couplings  to SM particles beyond gravitational interactions -- which ultimately  substantially reduces the experimental ability to determine their properties.

Nonetheless, cosmological probes have proven to be a comparatively powerful tool to test the phenomenological features of both of these species. Indeed, in the case of neutrinos the combination of Cosmic Microwave Background (CMB) and Baryon Acoustic Oscillations (BAO) data has been able to constrain the sum of neutrino masses to be $\Sigma m_\nu \leq 0.12$ eV (at 95\% CL) \cite{Aghanim2018PlanckVI}, which is further reduced to $\Sigma m_\nu \leq 0.09$ eV (at 95\% CL) when including \lya data \cite{Palanque-Delabrouille:2019iyz} (under the assumption of $\Lambda$CDM cosmology). Furthermore, the CMB not only provides one of the key pieces of evidence for the existence of DM, but also constrains its energy density \cite{Aghanim2018PlanckVI}, possible annihilation~\cite{Slatyer:2015jla, Aghanim2018PlanckVI}, decay~\cite{Poulin2016Fresh, Slatyer:2016qyl} and even interactions with other SM particles~\cite{AliHaimoud2015Constraints, Becker:2020hzj}.

Nevertheless, the emergence of various issues within the $\Lambda$CDM model, such as the $H_0$ and $\sigma_8$ tensions \cite{DiValentino2020CosmologyII, DiValentino2020CosmologyIII, DiValentino2021Realm, Schoneberg:2021qvd}, the $A_\mathrm{lens}$ and EDGES anomalies \cite{DiValentino2020CosmologyIV, Bowman:2018yin, Fraser:2018acy}, and the small-scale crisis \cite{Klypin:1999uc, 2011MNRAS.415L..40B, Tulin:2017ara}, has challenged the correctness of the standard cosmological model, and thereby also the robustness of the aforementioned constraints. Therefore, in particular in light of upcoming early- and late-universe missions such as CMB-S4~\cite{Abazajian2016CMB, Abazajian2019CMB}, the Simons Observatory~\cite{SimonsObservatory:2018koc}, and \text{EUCLID}~\cite{Amendola2012Cosmology}, which are expected to strengthen these constraints further (and potentially detect deviations from the standard predictions with high precision~\cite{Brinckmann2018Promising, Li:2018zdm, Gluscevic:2019yal,Cang:2020exa}), it becomes fundamental to test the validity of the bounds on $\Sigma m_\nu$ and on DM properties in scenarios beyond $\Lambda$CDM. This is especially significant for models that affect these quantities in the attempt to address one or more of the aforementioned issues.

In this regard, a particularly intriguing and long-standing possibility is that neutrinos and DM could themselves be coupled via a yet-undiscovered interaction channel. Many possible forms of the cross section (CS) ruling these interactions have been proposed in the literature, with ramifications on a variety of observables ranging from cosmology~\cite{2010PhRvD..81d3507S, Wilkinson:2013kia, Wilkinson:2014ksa, Campo:2017nwh, Bertoni:2014mva, DiValentino:2017oaw, Escudero:2018thh} to astrophysics~\cite{Kolb1987Supernova, Shoemaker:2015qul, deSalas:2016svi, Pandey:2018wvh, 1903.03302} and collider physics (see Sec. 3 of~\cite{Pandey:2018wvh}). 

Focusing in particular on the cosmological implications, DM-neutrino interactions (henceforth \text{iDM$\nu$}) are well known to affect both the CMB anisotropy power spectra and the late-time matter power spectrum (MPS) (see e.g.,~\cite{Wilkinson:2013kia, Wilkinson:2014ksa, Campo:2017nwh} for a complete review of the underlying physical effects). Of particular significance, due to the additional drag effect between the two species, these interacting models can erase structures on small scales, which results in an overall suppression of the MPS \cite{Wilkinson:2014ksa}, thus potentially solving the $\sigma_8$ tension~\cite{Mosbech:2020ahp}.

Given this impact on the MPS, observations of small-scale structure formation, such as the \lya forest flux spectrum, are crucial to constrain these models. Indeed, this probe measures the absorption lines in the spectra of quasars at redshifts $2\lesssim z \lesssim 6$ produced by clouds of neutral hydrogen in the intergalactic medium (IGM)~\cite{Viel:2005qj, Viel:2013fqw, Palanque-Delabrouille:2019iyz, Garzilli:2019qki} and, as such, has been shown to be highly sensitive to these scales~\cite{Ikeuchi1986, 10.1093/mnras/218.1.25P,Irsic:2017ixq}. It is, therefore, not surprising that the current most stringent bounds on iDM$\nu$ come from \lya data, which, under the assumption of a simple CS of the form ${\sigma_{\textrm{iDM}\nu}=\sigma_0\left(m_\textrm{DM}/\text{GeV}\right)}$\,, provide constraints of the order of ${\sigma_0\lesssim 10^{-33}}$~cm$^2$ for a 1~GeV DM particle~\cite{Wilkinson:2014ksa}, while CMB-driven constraints are orders of magnitude looser.

The analysis performed in~\cite{Wilkinson:2014ksa} was, however,  in the simplifying limit of massless neutrinos, while a more realistic treatment would require these to be massive. Nevertheless, although the effects of massive neutrinos on cosmology by themselves are well known~\cite{Lesgourgues:2013sjj}, their interactions with DM have only recently been implemented in a public cosmological Boltzmann solver in~\cite{Mosbech:2020ahp}, where iDM$\nu$ were constrained using only CMB and BAO data. Furthermore, the \lya results presented in~\cite{Wilkinson:2014ksa} were not directly derived from the data itself, but rather inferred from bounds on warm DM (WDM) previously derived in \cite{Viel:2013fqw}. However, as evident from Fig.~2 of \cite{Wilkinson:2014ksa}, the shape of the suppression induced on the MPS differs between the iDM$\nu$ and the WDM cases, such that a direct comparison to the data might deliver different results. With this in mind, here we update the analysis of \cite{Wilkinson:2014ksa} in two ways: by going beyond the assumption of massless neutrinos and with a direct comparison to the data.

Moreover, since \cite{Mosbech:2020ahp} found that, when considering only Planck and BAO data, iDM$\nu$ can successfully alleviate the $\sigma_8$ tension (often parametrised in terms of the parameter combination $S_8=\sigma_8\sqrt{\Omega_m/0.3}$), a further goal of this work will be to test whether iDM$\nu$ can still address the $\sigma_8$ tension after the inclusion of \lya data.

This paper is organised as follows. In Sec. \ref{sec: num} we describe the numerical setup and the cosmological probes used to constrain the aforementioned iDM$\nu$ model. In Sec. \ref{sec: res} we present the results obtained with the inclusion of \lya data (Sec. \ref{sec: res1}), together with additional discussions about the numerical caveats (Sec.~\ref{sec: res2}), a possible physical interpretation (Sec. \ref{sec: res4}), and a comparison with the literature (Sec. \ref{sec: res3}). We conclude in Sec.~\ref{sec: conc} with a summary and final thoughts. More details regarding the numerical results are provided in App.~\ref{app:params}.

\section{Numerical setup}\label{sec: num}
We use the publicly available version of the cosmology Boltzmann solver \textsc{class} \cite{Lesgourgues2011Cosmic, Blas2011Cosmic} developed and described in~\cite{Mosbech:2020ahp}, which can account for the presence of iDM$\nu$ (henceforth always implying massive neutrinos, also for $\Lambda$CDM), and allows to solve the Boltzmann equations for the massive neutrino species in the Newtonian gauge and in a momentum-dependent way\footnote{This is fundamental since massive neutrinos can transition from being ultra-relativistic to non-relativistic within cosmic times, and therefore different approximation schemes are required.}. Following this implementation, we assume the same constant CS as in~\cite{Wilkinson:2014ksa}, which is rescaled by introducing the parameter $u_{\text{iDM}\nu}=(\sigma_{\text{iDM}\nu}/\sigma_{\rm Th})\left(m_\textrm{DM}/100~\text{GeV}\right)^{-1}$, where $\sigma_{\rm Th}$ is the Thompson CS. As in \cite{Mosbech:2020ahp}, we further assume that the same interaction strength applies to all three neutrino species, that the total neutrino mass is negligible with respect to the DM mass, and that the total DM content of the universe is interacting.

To constrain iDM$\nu$, and to test their ability to address the aforementioned tension, we follow~\cite{Mosbech:2020ahp} in considering the Planck 2018 baseline dataset~\cite{Aghanim2018PlanckVI}, which includes temperature, polarisation and CMB lensing data, as well as BAO data gathered from 6dFGS at $z=0.106$ \cite{Beutler2011Galaxy}, SDSS from the MGS galaxy sample at $z=0.15$ \cite{Ross2014Clustering} and BOSS from the CMASS and LOWZ galaxy samples of SDSS-III DR12 at $z=0.2-0.75$ \cite{Alam2016Clustering}\footnote{Note that our choice of BAO datasets differs slightly from the one considered in \cite{Mosbech:2020ahp}. We check, however, that the constraints on the parameters of the model are unaffected by this choice.}. 
We will refer henceforth to the combination of these datasets as \textit{baseline}.

Furthermore, we extend the \textit{baseline} datasets to include also \lya data. One issue with \lya data, however, is that the scales involved lie in the non-linear regime, and therefore usually require computationally expensive hydrodynamical simulations. These simulations are very dependent on the underlying cosmological model, and are currently only available for a limited subclass of models \cite{Bolton:2016bfs}. In order to avoid the need for new hydrodynamical simulations, several workarounds have been proposed in the literature \cite{Viel:2005qj, Murgia:2017lwo, Archidiacono2019Constraining, Garny:2018byk}. Here we will make use of the approach developed in~\cite{Murgia:2017lwo}, which relies on an interpolation within a pre-computed grid of simulations (see Sec. III of the reference as well as~\cite{Archidiacono2019Constraining} for the technical details of the simulations, including the handling of the astrophysical parameters) and considers data from the HIRES/MIKE samples of quasar spectra~\cite{Viel:2013fqw}, which cover the redshift range $z = 4.2 - 5.4$. This approach was later integrated by \cite{Archidiacono2019Constraining} into a full likelihood for the parameter extraction code \textsc{MontePython} \cite{Audren2013Conservative, Brinckmann2018MontePython} (see Sec. 3.1 of \cite{Archidiacono2019Constraining} for more information on the numerical implementation). 

As described in~\cite{Murgia:2017lwo, Archidiacono2019Constraining}, however, the simulations that constitute the grid assume massless neutrinos, so that some care has to be taken before making use of this likelihood for the cosmological model considered here. Nonetheless, as explained in the introduction of~\cite{Palanque-Delabrouille:2015pga}, the ratio of the MPSs between massive and massless neutrinos is almost flat at \lya scales (see Fig.~1 of the reference). For this reason, \lya data are only weakly sensitive to the difference between massive and massless neutrinos, which can be almost completely absorbed by a change in the MPS amplitude and tilt (which could be parameterised via e.g., $\sigma_8$, $\Omega_m$ and $n_s$ or $n_{\rm eff}$ -- the slope of the MPS at \lya scales), as can be seen in e.g., Fig. 2 of \cite{Pedersen:2019ieb} and related text. Therefore, as the grid of simulations fully allows for variations within these quantities ($\sigma_8$ and $n_{\rm eff}$), the massive case is already intrinsically accounted for.\footnote{More explicitly, the effects of $\sum m_\nu$ and the right combination of $\{\sigma_8,n_{\rm eff}\}$ lead to practically the same MPS suppression and hence to the same \lya flux. Therefore, \lya data are largely unable to distinguish between the introduction of massive neutrinos and a corresponding shift in $\{\sigma_8,n_{\rm eff}\}$. This implies that for a given value of $\sum m_\nu$ the likelihood derives the corresponding values of $\{\sigma_8,n_{\rm eff}\}$, passes them as an input to the grid interpolator, and the correct flux for that given $\sum m_\nu$ is generated. We note, however, that this discussion is only true up to a minor redshift dependence of the suppression, which is well below the sensitivity of the data used, and can therefore be safely neglected here.}

We can then turn our attention to iDM$\nu$. In this scenario we consider an extension of the $\Lambda$CDM model with $\{\omega_{\rm b}, \omega_{\rm dm}, H_0, \ln(10^{10}A_s), n_s, \tau_{\rm reio}\}+\{\Sigma m_\nu,u_{\rm iDM\nu}$\}. As in \cite{Mosbech:2020ahp}, we impose a lower limit on the sum of neutrino masses of the form $\Sigma m_\nu>0.06$ eV, which corresponds to the minimal theoretical value assuming normal hierarchy (we have checked that this prior does not impact our result for the CS). We also adopt the same fluid approximation described in \cite{Lesgourgues_2011} and App. B of \cite{Mosbech:2020ahp}, with the knowledge that the error it introduces (even at \lya scales, as we check explicitly) is much smaller than the uncertainties on the data we employ \cite{Viel:2013fqw}.

\section{Results}\label{sec: res}
\subsection{Main results}\label{sec: res1}
\begin{figure}[t]
	\centering
	\includegraphics[width=0.95\columnwidth]{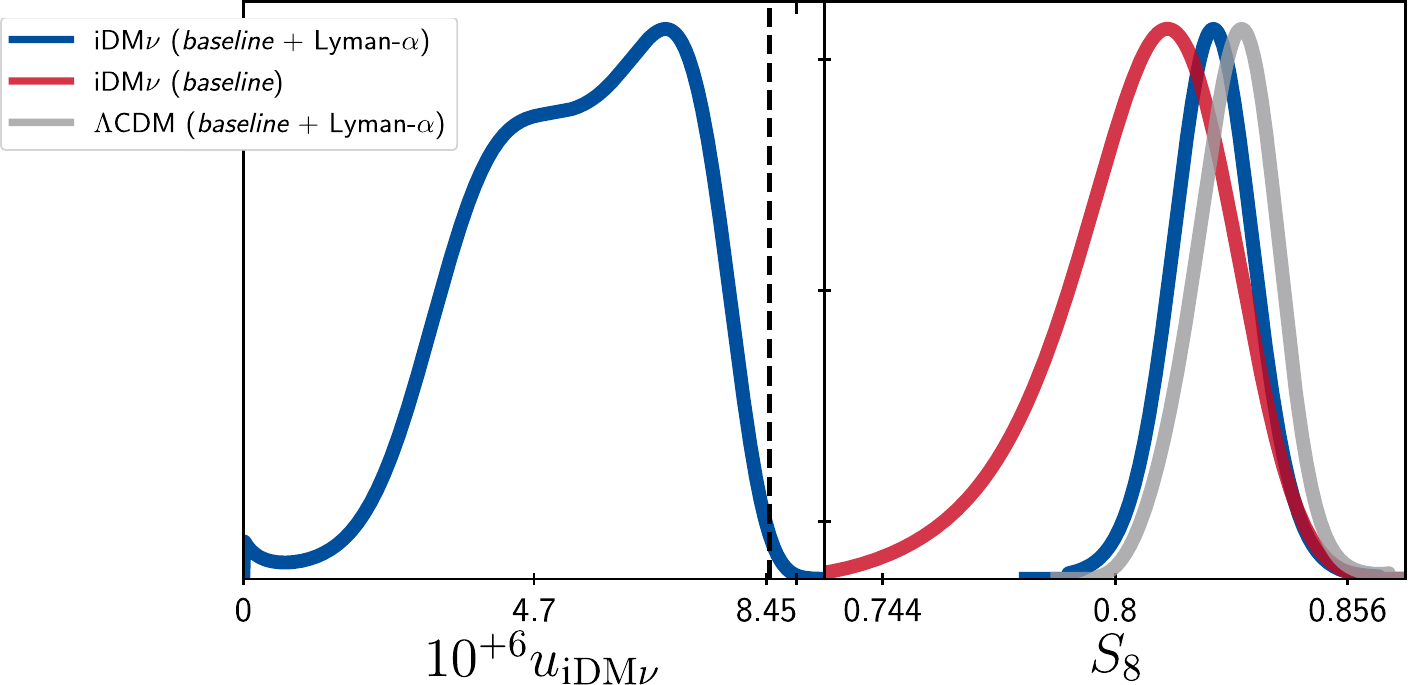}
	\caption{1D distributions of the $u_{\rm iDM\nu}$ (left) and $S_8$ (right) parameters for the different models and datasets considered. The \textit{baseline}-only contours for $u_{\rm iDM\nu}$ are not displayed, as they are two orders of magnitude wider (and can already be found in \cite{Mosbech:2020ahp}). The dashed black line shows the boundary of applicability of the \lya likelihood.}
	\label{fig: MCMC_res}
\end{figure}

In order to test our pipeline against the results found in~\cite{Mosbech:2020ahp}, we first perform a Markov Chain Monte Carlo (MCMC) run considering only the \textit{baseline} data, obtaining results largely compatible with the reference. Then, analogously to what was done in \cite{Archidiacono2019Constraining}, we run two MCMC simulations. The first one (henceforth referred to as \tquote{no-data run}) only considers the so-called \tquote{applicability checks} listed in \cite{Archidiacono2019Constraining} (i.e., no actual \lya data are accounted for) and only serves to define the region of parameter space where the \lya likelihood is applicable. The second (henceforth referred to as \tquote{data run}) includes also the information from the \lya data. Starting by the no-data run, we find that all values of the interaction strength $u_{\rm iDM\nu}$ roughly below $8.5\times10^{-6}$ can be probed by the likelihood (see the dashed vertical line in Fig.~\ref{fig: MCMC_res}) with an almost flat distribution. We then perform the full MCMC run for iDM$\nu$ including \lya data. In Fig. \ref{fig: MCMC_res} we display the posterior distributions for the most relevant parameters (with a full list of constraints provided in App.~\ref{app:params}).

The results for the CS, surprisingly, hint to a departure from the $\Lambda$CDM prediction with ${u_{\rm iDM\nu}=5.5_{-1.1}^{+2.6}\times 10^{-6}}$ (at 1$\sigma$), implying a deviation from $\Lambda$CDM of about  3$\sigma$ (assuming the distribution to be Gaussian and considering the $\Delta \chi^2$ discussed below -- see also the detailed $\chi^2$ breakdown in App.~\ref{app:params}). Importantly, this is not indicative of the overall preference for this model over $\Lambda$CDM, but only reflects the fact that the values that can best correct the tilt of the \lya flux for this particular model lie roughly 3$\sigma$ away from zero.

In addition to this, we also find that, after the inclusion of \lya data, the iDM$\nu$ model can no longer solve the $\sigma_8$ (or $S_8$) tension, and we observe no change in the upper bound on $\Sigma m_\nu$ with respect~to~\cite{Mosbech:2020ahp} (see Tab.~\ref{tab:full} in App.~\ref{app:params}). The reason for the former result is that the degeneracy between the $\sigma_8$ and $u_{\rm iDM\nu}$ parameters -- which was allowing the model to solve the tension when only accounting for the \textit{baseline} data sets (see Fig. 4 of \cite{Mosbech:2020ahp}) -- is significantly broken when accounting for \lya data, which improve the constraints on $u_{\rm iDM\nu}$ by two orders of magnitude.

Since the found value of $u_{\rm iDM\nu}$ is almost two orders of magnitude smaller than the upper bound reported for the \textit{baseline} datasets \cite{Mosbech:2020ahp}, it must be primarily driven by the inclusion of \lya data (also confirmed on the basis of a $\Delta\chi^2$ analysis -- where \lya data prefer iDM$\nu$ with $\Delta\chi^2\simeq-7.5$, and the \textit{baseline} penalise it by roughly the same amount, see Tab. \ref{tab:chi2} in App. \ref{app:params}) and it can mainly have two possible origins: numerical or physical\footnote{The presence of possible yet-unknown systematics in the considered data sets might also be a possibility to explain the presence of the preference. However, we will not consider this avenue here and instead refer the interested reader to the data-release papers for dedicated discussions.}. 

\subsection{Numerical caveats}\label{sec: res2}
From the numerical side, although the distribution lies relatively close to the edge of the allowed region, we observe no sharp cutoff nor hint to the possibility that the position of the peak might be method driven. Moreover, as explained before, we have also carefully tested the likelihood and the Boltzmann solver, finding very good agreement with the literature.
\begin{figure*}[t]
	\centering
	\includegraphics[width=\columnwidth]{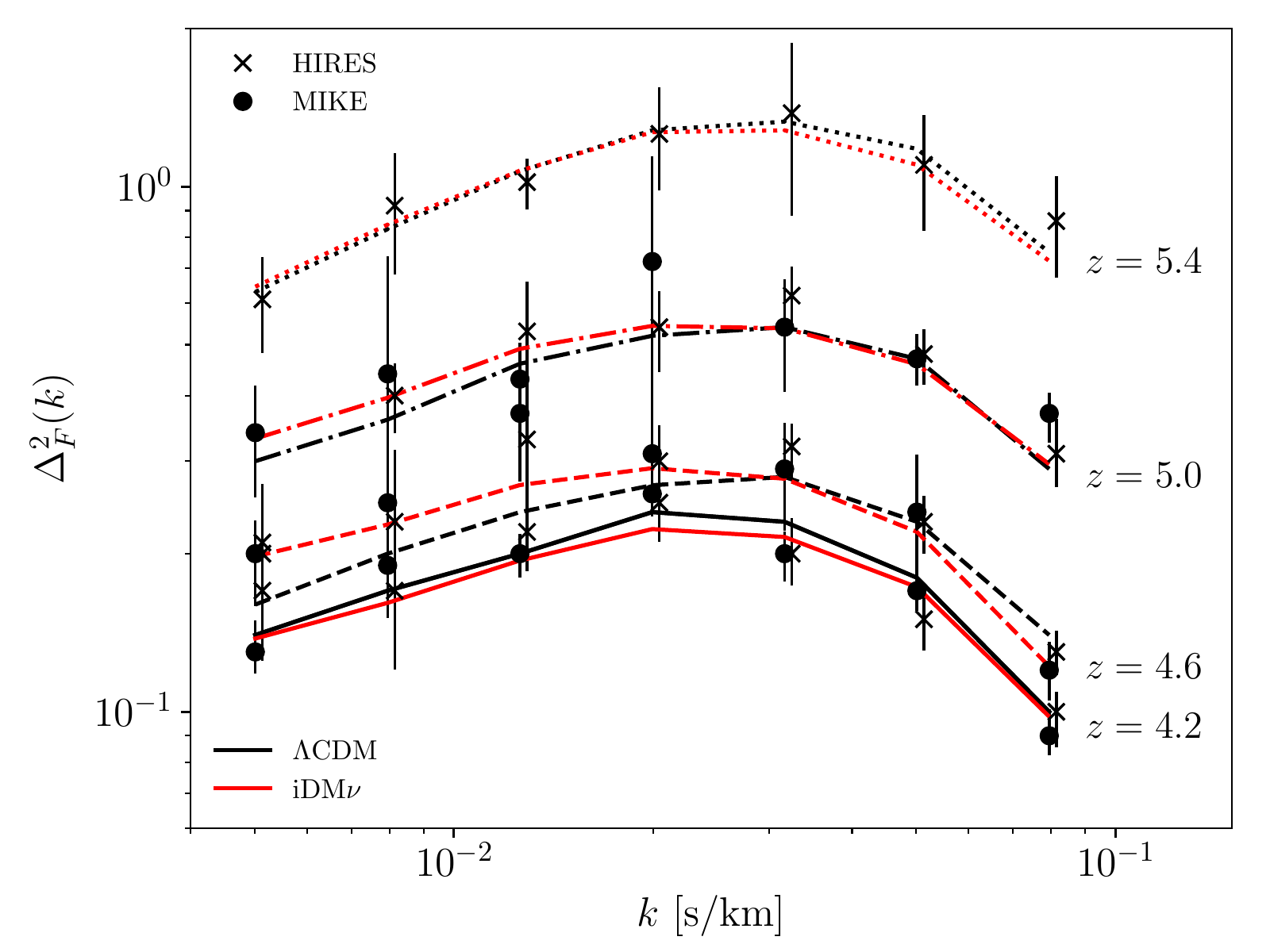}	\includegraphics[width=\columnwidth]{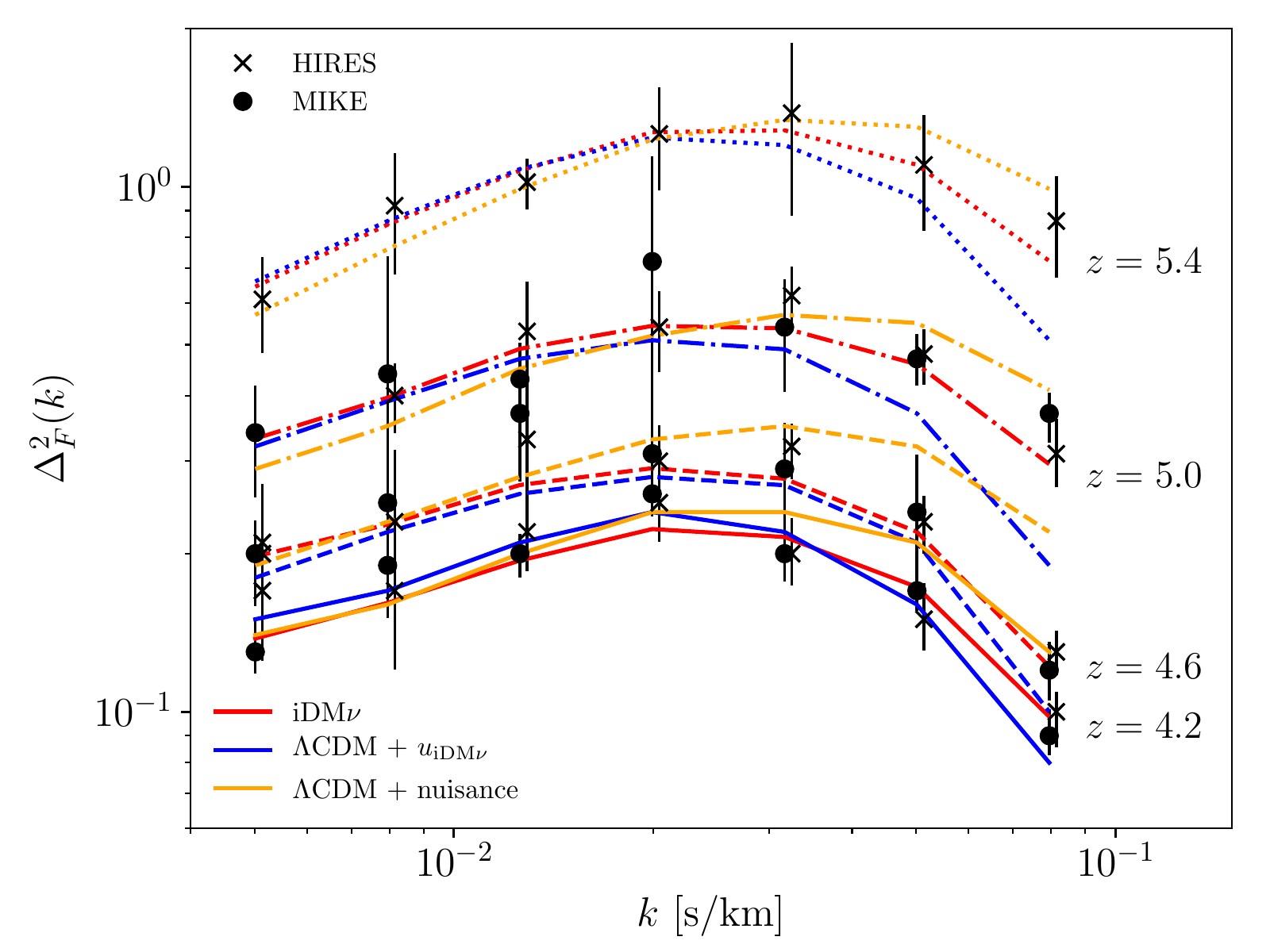}
	\caption{\textit{Left panel:} Comparison between the $\Lambda$CDM (black) and iDM$\nu$ (red) best-fitting models. \textit{Right panel:} Comparison between two models with $\Lambda$CDM best-fitting values for all parameters except for $u_{\rm iDM\nu}$ (blue) and the \lya nuisance parameters (orange), for which we use the iDM$\nu$ best-fitting value(s), and the iDM$\nu$ model (red, same as in left panel). In both panels, crosses and dots represent HIRES/MIKE data.}
	\label{fig: test_flux}
\end{figure*}

Of note, however, is the potential role of $n_{\rm eff}$, which is derived from the initial parameters and used as an input to infer the \lya flux form the pre-computed grid of simulations. This approach, as described in~\cite{Murgia:2018now}, has been tested in the range $[-2.3474, -2.2674]$ and only extrapolated to the wider range $[-2.6, -2.0]$, with a warning of possible extrapolation difficulties beyond the tested region.~Therefore, since for the iDM$\nu$ case the aforementioned value of $u_{\rm iDM\nu}$ corresponds to\footnote{As $n_{\rm eff}$ is by the definition related to the tilt of the MPS at Lyman-$\alpha$ scales, which is in turn directly related to the interaction strength (that determines how suppressed the spectrum is at a given scale), a strong degeneracy between the two parameters is to be fully expected.} ${n_{\rm eff}=-2.507_{-0.087}^{+0.038}}$ (see Fig. \ref{fig: MCMC_res_nuisance} in App. \ref{app:params}), future tests of the solidity of the extrapolation scheme will be needed. On the other hand, however, in \cite{Murgia:2018now} it was argued that the cosmological predictions are not impacted by these possible deviations in $n_{\rm eff}$, although that might be have been due to absence of correlations between $n_{\rm eff}$ and the other parameters of the models considered in the reference. 

Furthermore, we also observe a similar occurrence with some of the parameters describing the IGM temperature evolution (in particular $T_{0,a}$, see Tab.~\ref{tab:full} and Fig.~\ref{fig: MCMC_res_nuisance} in App.~\ref{app:params}). As discussed in detail in the following section, within the iDM$\nu$ model a colder thermal history is preferred with respect to what is expected within $\Lambda$CDM (compare e.g., Fig. 1 of~\cite{Irsic:2017ixq} and Fig.~E10 of \cite{Boera:2018vzq} to Tab.~\ref{tab:full} in App.~\ref{app:params}), differing at the $1-2\sigma$ level. On the one hand, one could argue, as done in e.g.,~\cite{Garzilli:2018jqh,Garzilli:2019qki}, that this apparent contradiction with the literature might not be problematic, as the discussions carried out in \cite{Irsic:2017ixq, Boera:2018vzq} (and several other references focusing e.g., on the relation to reionization \cite{Gaikwad:2020art}) were based on \lya data with the assumption of $\Lambda$CDM, and as such might not be applicable to scenarios with large, small-scale suppressions such as WDM (see e.g., Fig. 4 of~\cite{Garzilli:2019qki}) and iDM$\nu$. However, in our case this colder thermal history could introduce numerical artefacts in the aforementioned results, since the preferred iDM$\nu$ values lie below the tested region of parameter space covered by the grid of simulations.

In other words, for both $n_{\rm eff}$ and some of the parameters ruling the thermal history of the IGM, in our analysis we are pushing the \lya likelihood discussed in Sec.~\ref{sec: num} to its boundaries, although still formally within the limits considered reliable by the developers~\cite{Murgia:2018now, Archidiacono2019Constraining}. Moreover, we add that while a more thorough investigation of both of these aspects of the analysis will certainly be crucial in future work, the fact that $\Lambda$CDM-like values of $n_{\rm eff}$ and $T_{0,a}$ are disfavoured by \lya data with respect to the iDM$\nu$ values is, in principle, completely independent of the extrapolation details within the grid, and still supports the idea of a non-zero interaction strength.

\subsection{Physical interpretation}\label{sec: res4}
Having discussed the numerical caveats, we can now turn our attention to the exciting possibility that this preference might be of physical nature. The two main questions that need to be addressed in this case are \textit{\text{What effect(s) could drive such a preference?}} and \textit{Why was it not found before for other models?}\footnote{Although it can be argued that WDM or similar models are not necessarily excluded by \lya data \cite{Garzilli:2018jqh}, a \textit{preference} was not found before and hence the aforementioned question is still valid.}

One possible avenue to address the first question stems from a more careful analysis of the \lya flux (which is the true observable) and in particular from e.g., Fig.~13 of \cite{Viel:2013fqw}, where the HIRES/MIKE data used to derive our bounds are shown. There it is clear that, although $\Lambda$CDM provides a very good fit to the data, at large scales (below $k\simeq 0.02$~s/km) deviations start to grow between its prediction and the data points. Also, at small scales in particular HIRES (Fig.~12 of \cite{Viel:2013fqw}) leaves room for an additional suppression of the flux power spectrum that the $\Lambda$CDM prediction cannot fully accommodate. This means that a model with an overall additional tilt of the \lya flux power spectrum enhancing large scales and suppressing small scales would fit the data better than $\Lambda$CDM.

We graphically investigate this possibility for the iDM$\nu$ model in the left panel of Fig.~\ref{fig: test_flux}. There we compare the \lya flux power spectrum predicted by our $\Lambda$CDM best-fitting values (black) and the iDM$\nu$ counterpart (red), with the MIKE/HIRES data points shown for reference. As anticipated, from the figure it is clear that iDM$\nu$ are correctly adjusting the overall tilt of the spectrum, while the $\Lambda$CDM prediction (whose $n_s$ and $\Omega_m$ values are most significantly driven by the \textit{baseline} datasets, as was the case in \cite{Viel:2013fqw} due to their priors) is more strongly deviating from the data points. From the figure it also becomes clear that the majority of the $\chi^2$ improvement of iDM$\nu$ over $\Lambda$CDM is driven by the intermediate redshift bins ($z=4.5$ and $z=5.0$) and modes below $k\simeq 0.03$, with a minor contribution from the small scales of the redshift bin~$z=4.6$.

A qualitative explanation for why the iDM$\nu$ model can perform better than $\Lambda$CDM is displayed in the right panel of the same figure. There we show the $\Lambda$CDM prediction but with $u_{\rm iDM\nu}$ (blue) and the \lya nuisance parameters (orange) set to their iDM$\nu$ best-fitting values. As it turns out, the individual contribution from iDM$\nu$ is indeed sharply cutting off the spectrum at small scales (as in the MPS, see Fig. \ref{fig: test_pk}), but the nuisance parameters can largely compensate for this suppression (see \cite{Garzilli:2018jqh} for related discussions) without significantly spoiling the additional tilt introduced by iDM$\nu$ at large scales.

Moreover, very recently the cosmological implications of complementary \lya data gathered by SDSS DR14 BOSS and eBOSS \cite{2013AJ....145...10D, Chabanier:2018rga} (probing larger scales, up to $k= 0.02$ s/km, and lower redshifts than MIKE/HIRES) have been accurately investigated in \cite{Palanque-Delabrouille:2019iyz}. There, the authors point out a tension in the determination of the scalar spectral index $n_s$ (or equivalently in the matter density parameter $\Omega_m$) between the \textit{baseline} dataset and their \lya data, confirming a similar finding previously discussed in \cite{Palanque-Delabrouille:2015pga} and based on SDSS DR9 BOSS data. Interestingly, in \cite{Palanque-Delabrouille:2019iyz} they performed an analysis with different values of $n_s$ for the \textit{baseline} and the \lya likelihoods, finding that if the slope of the  MPS at \lya scales is sharpened (such that $n_s=0.941$ at \lya scales, against the $n_s=0.967$ that they obtain for Planck) the compatibility of the two probes is greatly increased. 

\begin{figure}[t]
	\centering
	\includegraphics[width=\columnwidth]{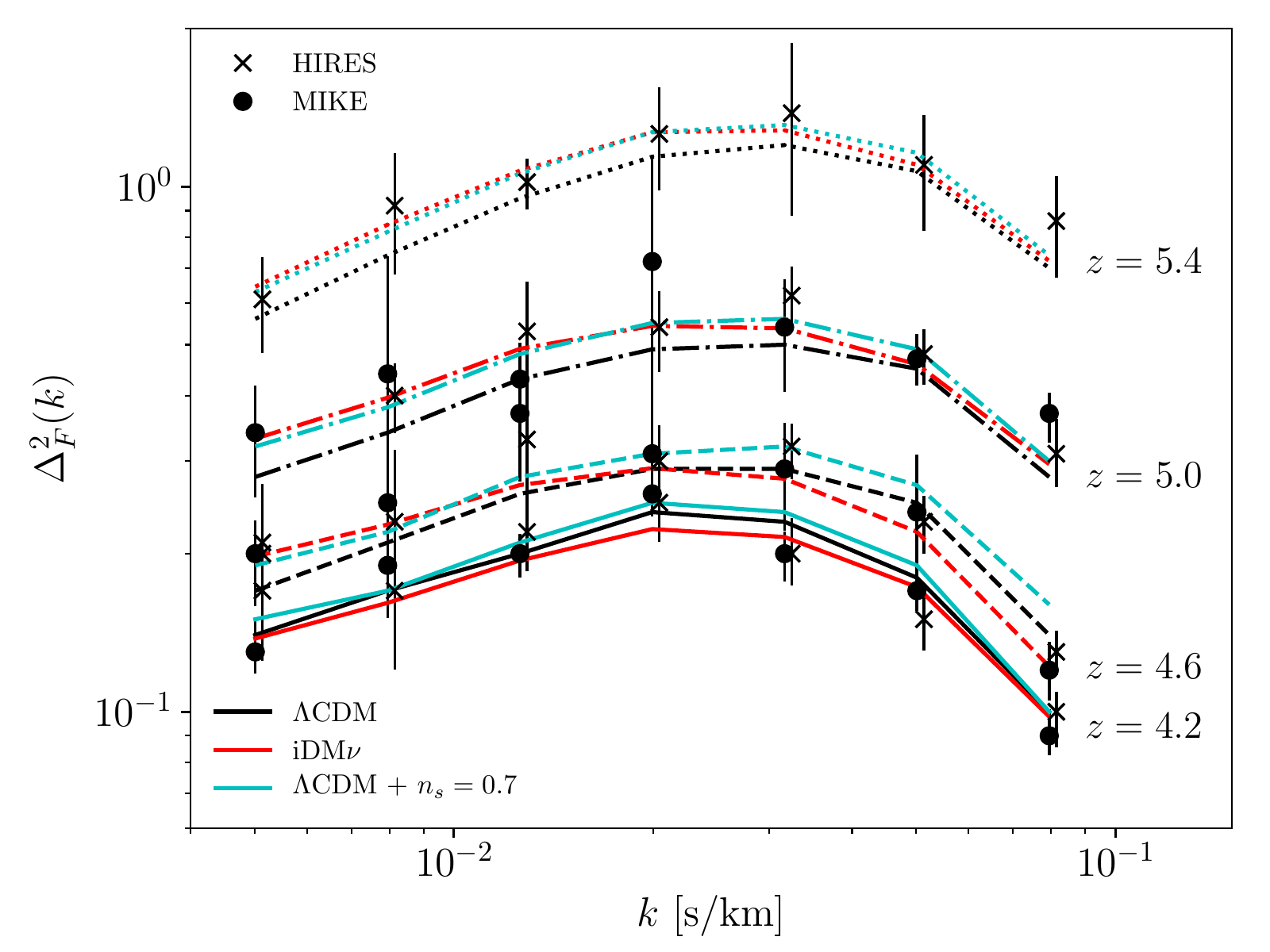}
	\caption{Same as in the left panel of Fig.~\ref{fig: test_flux}, but compared to the $\Lambda$CDM model with $n_s=0.7$ (light blue).}
	\label{fig: test_flux_2}
\end{figure}

This confirms our qualitative discussion based on Fig.~13 of \cite{Viel:2013fqw}\footnote{Although such an analysis has not yet been performed in the context of the MIKE/HIRES data  used here, we can still largely extend its conclusions to our discussion. In fact, this preference for a more pronounced tilt of the MPS at \lya scales was already present in SDSS DR9 BOSS data \cite{Palanque-Delabrouille:2015pga}, which \cite{Viel:2013fqw} found to be very compatible with MIKE/HIRES data.}, and to show that this is indeed the case for our iDM$\nu$ model we plot in Fig. \ref{fig: test_flux_2} a model with the same $\Lambda$CDM best-fits as for the black curves but with an $n_s$ value arbitrarily decreased to $n_s=0.7$ (light blue). This value is set such that it roughly matches the iDM$\nu$ predictions at large scales for graphical purposes and, although being clearly unrealistic, would also incorporate eventual changes due to a decreased $\Omega_m$ value (which is also required to address the tension pointed out in \cite{Palanque-Delabrouille:2019iyz}). What one can then see in the figure is that the interacting model can indeed mimic the effect of a decreased $n_s$ (and $\Omega_m$) at large scales and in particular at the scales probed by SDSS DR14 BOSS and eBOSS, which were the basis for the analysis of \cite{Palanque-Delabrouille:2019iyz}. While this is an important proof of principle, a more thorough analysis comparing directly to those datasets is left for future work.

As a further remark, note that damped \lya absorbers (DLAs), which are currently not included in our analysis, could also play a role in the cosmological parameter estimation \cite{Rogers:2017bmq}. Nevertheless, as found in \cite{Palanque-Delabrouille:2019iyz}, at least in the context of BOSS data the presence and strength of the tension in $n_s$ is unaffected by this contribution.

\subsection{Comparison with the literature}\label{sec: res3}
It seems, therefore, that the ability of the model to mimic a lower $n_s$ value at large scales and to introduce a (overall moderate) suppression of the flux at small scales are indeed the key ingredients driving the preference for iDM$\nu$ over $\Lambda$CDM. It is, however, important to consider why other well-known models introducing a small-scale suppression of the MPS and of the flux, such as WDM, DM-dark radiation (iDMDR) and DM-massless neutrinos interactions, do not deliver the same preference, and thereby address the second question mentioned before.

Starting with WDM, one can focus on Figs.~11-13 of \cite{Viel:2013fqw} and see that the effect of WDM on the \lya flux (green versus red curves therein) is very distinct from what was discussed here in the context of iDM$\nu$. Indeed, WDM almost only suppresses the spectrum\footnote{Note that from Fig. 3 of \cite{Viel:2013fqw} it is clear that the WDM scenario also enhances the spectrum at large scales to a certain extent. However, at the level of the \textit{dimensionless} flux power spectrum $\Delta^2_F$, which is instead shown in Fig. \ref{fig: test_flux} and Figs. 11-13 of the reference, these deviations at large scales significantly reduce.} at small scales without any significant additional enhancement at large scales, which means that the overall tilt of the flux does not change. In other words, WDM does not seem to reproduce the same impact on the flux power spectrum at large scales as a change in $n_s$ (see Fig.~\ref{fig: test_flux_2}), which, as argued above, is necessary in order to reconcile the prediction of the model with the data. Focusing then on the MPS, one can generalise this conclusion in terms of how sharply a given model is suppressing the spectrum: models with too sharp suppressions are intrinsically unable to emulate a change in the tilt, while models introducing softer suppressions might achieve this behaviour. Indeed, WDM is more sharply suppressing the MPS with respect to our iDM$\nu$ model, as clear from e.g., Fig. 2 of \cite{Wilkinson:2013kia}.

We explicitly check that this interpretation is also consistent with the case of iDMDR analysed in \cite{Archidiacono2019Constraining}, where the same numerical pipeline was used as in our work and no preference for the interacting models was found. There various interacting models were considered with dependence on the scaling index $n$ of the comoving interaction rate (assumed to be equal to 0, 2 or 4 in the reference), the temperature ratio between DR and photons $\xi$, and the interaction strength $a_{\rm dark}$ (see the reference for more details on the notation). As clear from Fig.~1 of~\cite{Archidiacono2019Constraining}, the shape of the suppressions can range from being very sharp in the the $n=4$ case to very soft in the $n=0$ scenario.
	
\begin{figure}[t]
	\centering
	\includegraphics[width=\columnwidth]{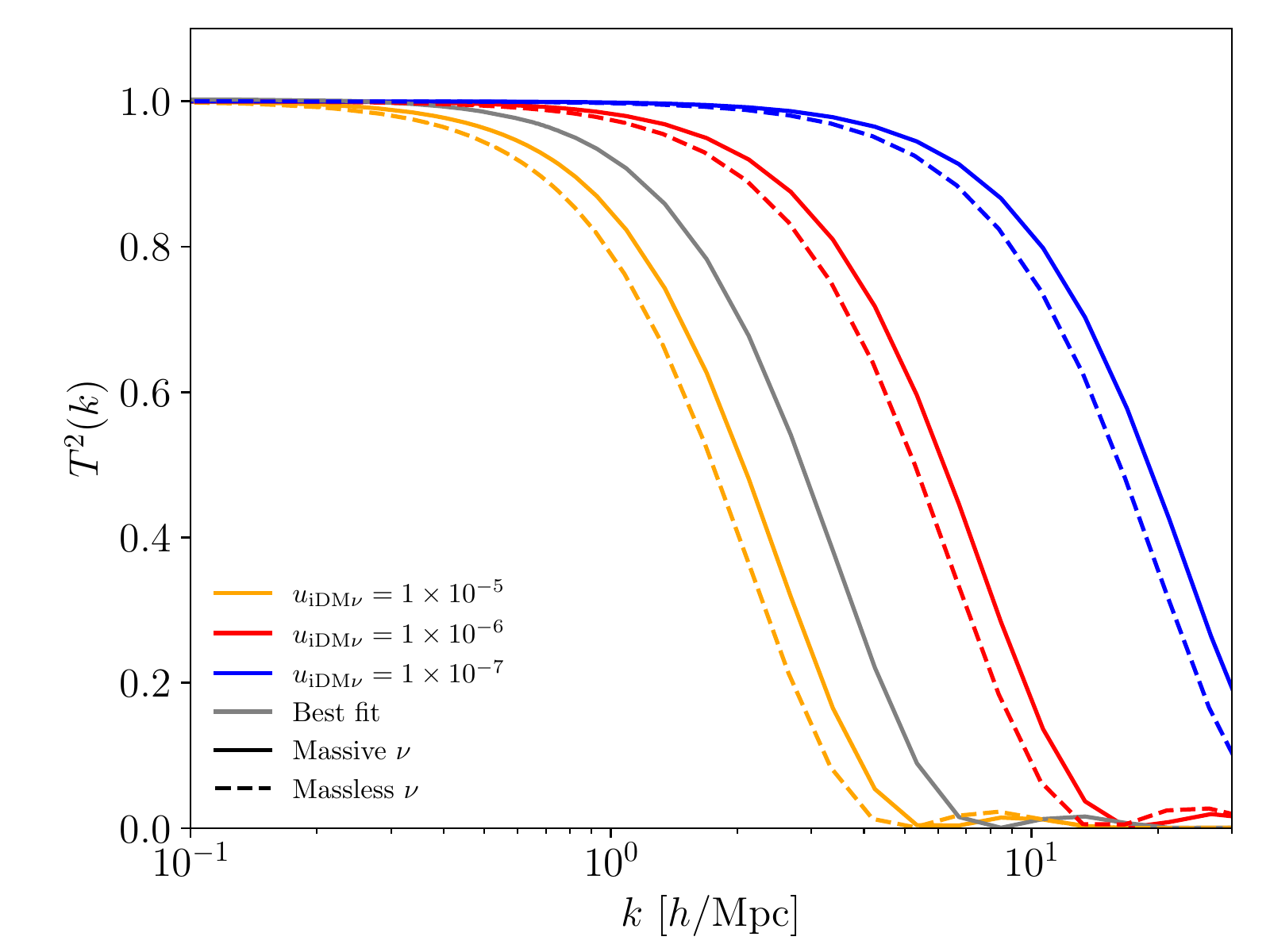}
	\caption{Representative examples of transfer functions of the iDM$\nu$ model (solid for massive neutrinos and dashed for massless neutrinos) for different values of the interaction strength $u_{\rm iDM\nu}$ as well as for the best-fitting value reported in Tab.~\ref{tab:full}.}
	\label{fig: test_pk}
\end{figure}

In the $n=0$ case, both $\xi$ and $a_{\rm dark}$ affect the position in $k$ and the shape (tilt) of the MPS suppression. However, based on the results of~\cite{Archidiacono2019Constraining} we find that in this scenario no combination of parameters allowed by Planck+BAO data can predict a full suppression of the MPS at $k$ values in the range between $1-10$ $h$/Mpc, i.e., where our best-fitting curve is found (see Fig. \ref{fig: test_pk}). Indeed, as already shown there, the models tend to induce a plateau-like suppression, which the \lya likelihood is not equipped to deal with. We therefore conclude that because the model predicts a too soft MPS suppression, the region of parameter space in the $n=0$ case that might resemble our preferred model is already excluded by early-time probes, thus explaining why no preference was found in~\cite{Archidiacono2019Constraining} for this model. 
	
For the $n=2,4$ cases, $\xi$ influences both the position in $k$ and the tilt of the suppression, while $a_{\rm dark}$ only affects its position in $k$. In both of these iDMDR models we find parameter combinations which predict a MPS suppression at the same scales as our best-fitting model (regardless of its tilt) along the whole degeneracy lines between $\xi$ and $a_{\rm dark}$, very close to the boundary of the no-data runs performed in the reference (similar to what we found in the iDM$\nu$ model, see Fig. \ref{fig: MCMC_res}). For instance, in the $n=2(4)$ case we are able to reproduce the exact same best-fit curve reported in Fig.  \ref{fig: test_pk} for the iDM$\nu$ scenario with the combination $\xi=0.3(0.05)$ and $\log_{10}(a_{\rm dark}/[\text{Mpc}^{-1}])=4(21)$. We also checked explicitly that with these parameter choices we obtain the same \lya $\Delta\chi^2$ improvement over $\Lambda$CDM as in iDM$\nu$, confirming that a better fit than $\Lambda$CDM exists in the iDMDR models as well. We conclude, therefore, that the absence of a preference in \cite{Archidiacono2019Constraining} has to trace back to the choice of logarithmic prior used therein. Indeed, while on the one hand a linear exploration of the parameter space is inadequate in the iDMDR scenario due to the wide range of values covered by $a_{\rm dark}$, on the other hand the logarithmic prior also reduces the sensitivity of the MCMC exploration to the narrow region of parameter space close to the edge of the distribution that would have been preferred by the data. Although currently inconclusive, a more in-depth discussion about a potential preference for iDMDR models is left for future work.

Moreover, the bounds derived for WDM have often been adapted and applied to a variety of other models, such as in the aforementioned interacting DM-massless neutrinos case \cite{Wilkinson:2014ksa}. It is, therefore, possible that other models that would have also been preferred by \lya data have not been identified because of the indirect comparison with the data. We confirm this precisely for the case of DM-massless neutrinos interactions. 
Indeed, by comparing the transfer functions for the massive and massless cases when assuming the same interaction strength, as shown in Fig.~\ref{fig: test_pk}, one can see that the shape of the suppression induced by the two scenarios is identical, and that the two curves are just offset horizontally by a factor of about 30\%. This difference can be explained by the fact that in the massive case the neutrinos eventually become non-relativistic during cosmic times, thereby contributing to structure formation and causing a smaller suppression of the MPS for the same interaction strength. Therefore, since, as argued above, the key property a model should have in order to correctly adjust the tilt of the flux power spectrum relies on the tilt it imprints on the MPS, and since the shape of the massless case is the same as in the massive case, we expect to find the same preference for the massless case but with a distribution shifted by 30\% with respect to the massive scenario. This is precisely what we find: $u_{\rm iDM\nu}=3.84_{-0.69}^{+1.8}\times 10^{-6}$ (at 1$\sigma$), with a similar distribution as the one shown in Fig. \ref{fig: MCMC_res}. As a remark, this consistency (i.e., the fact that the offset in the MPS between massless and massive cases is the same as the one in the mean values of the interaction strengths $u_{\rm iDM\nu}$) also confirms that the likelihood can indeed be safely applied to the massive case.

\section{Conclusions}\label{sec: conc}
In this work we considered a scenario allowing for (scattering) interactions between DM and massive neutrinos. We focused in particular on their impact on the \lya flux power spectrum, and constrained them with \lya data from MIKE and HIRES. Intriguingly, we found a preference for an interaction strength about $3\sigma$ away from zero (and that the model is unable to address the $S_8$ tension, contrary to what was previously thought). We have argued that this might be due to the ability of the model to add an overall additional tilt to the \lya flux power spectrum, simultaneously enhancing large scales and suppressing small scales more than $\Lambda$CDM.

Of course, such conclusions will have to be rigorously tested in future works. In particular, investigating the role of complementary \lya data will be crucial to test the validity of our results, as well as that of alternative constraints, such as the ones presented in~\cite{Escudero:2018thh}. Dedicated numerical simulations focusing on the role of the astrophysical nuisance parameters and $n_{\rm eff}$ will also be fundamental. 

We emphasize, however, that the fact that a better fit to the data than $\Lambda$CDM exists is independent of possible numerical artefacts and not intrinsic to the model itself. Indeed, the same result could be reproduced by any model that induces a suppression of the MPS with the correct softness and at the correct scales. In this light, the study carried out here goes beyond DM-neutrino interactions, highlighting what a model would need in order to successfully fit the \lya data considered here, and to potentially alleviate the known tension between \lya and early-universe probes in the determination of the tilt of the MPS at \lya scales.

\newpage

\section*{Acknowledgements} 
We thank T. Hambye, J. Lesgourgues, L. Lopez-Honorez, R. Murgia, N. Sch\"oneberg, and M. Viel for very useful discussions. ML is supported by an F.R.S.-FNRS fellowship, by the \tquote{Probing dark  matter with neutrinos} ULB-ARC convention and by the IISN convention 4.4503.15. DH was supported by the FNRS research grant number~\mbox{F.4520.19}, and by the Academy of Finland grant no. 328958. Computational resources have been provided by the CÉCI, funded by the F.R.S.-FNRS under Grant No. 2.5020.11 and by the Walloon Region. 

This project was carried out while both authors were grieving the loss of a parent. We dedicate this paper to their memory and their constant support and love.

\appendix

\section{Supplementary material on MCMC results}\label{app:params}
In this Appendix we provide a full list of cosmological and astrophysical constraints derived from our MCMCs including \lya data (Tab. \ref{tab:full}) as well as a complete breakdown of the individual $\chi^2$ values per experiment (Tab. \ref{tab:chi2}). The results presented in Tab.  \ref{tab:full} are to be compared to those listed in Tab. 3 of \cite{Mosbech:2020ahp} for dataset combinations without Lyman-$\alpha$. For the notation regarding the astrophysical (nuisance) parameters we refer the reader to \cite{Murgia:2018now}.

\begin{table}[t!]
	\def\arraystretch{1.2}
	\scalebox{0.9}{
		\begin{tabular}{|l|c|c|} 
			\hline
			& $\Lambda$CDM & iDM$\nu$ \\
			\hline
			$H_0$ [km/s/Mpc] & $67.75(68.12)_{-0.45}^{+0.52}$  & $67.43(67.66)_{-0.48}^{+0.48}$ \\
			$100~\omega_b$ & $2.240(2.249)_{-0.012}^{+0.014}$  & $2.241(2.237)_{-0.014}^{+0.013}$ \\
			$\omega_{\rm cdm}$ & $0.1194(0.1193)_{-0.00095}^{+0.00088}$ & $0.1192(0.1193)_{-0.00089}^{+0.001}$ \\
			$10^{9}A_s$ & $2.107(2.107)_{-0.032}^{+0.026}$ & $2.115(2.156)_{-0.034}^{+0.028}$ \\
			$n_s$ &  $0.9655(0.9669)_{-0.0037}^{+0.0036}$ &  $0.9662(0.9679)_{-0.0038}^{+0.0039}$ \\
			$\tau_{\rm reio}$ & $0.05639(0.05776)_{-0.0082}^{+0.0063}$ & $0.05867(0.06881)_{-0.0078}^{+0.0073}$ \\
			$\sum m_\nu$ [eV] & $<0.12$ & $<0.15$ \\
			$10^{+6}~u_{\text{iDM}\nu}$ & - & $5.45(5.14)_{-1.0}^{+2.5}$ \\
			\hline
			$z_{\rm reio}$ & $7.86(7.98)_{-0.79}^{+0.68}$ & $8.09(9.09)_{-0.75}^{+0.72}$ \\
			$n_{\rm eff}$ & $-2.3052(-2.3038)_{-0.0032}^{+0.0031}$ & $-2.507(-2.495)_{-0.087}^{+0.038}$ \\
			$\sigma_8$ & $0.8151(0.8218)_{-0.0077}^{+0.0097}$ & $0.8054(0.8212)_{-0.0072}^{+0.0091}$ \\
			$S_8$ & $0.829(0.830)_{-0.011}^{+0.010}$ & $0.824(0.836)_{-0.011}^{+0.011}$ \\
			\hline
			$10^4 ~T_{0,a}$ [K] & $1.15(0.89)_{-0.42}^{+0.44}$ & $0.34(0.48)_{-0.13}^{+0.21}$ \\
			$T_{0,s}$ & $<1.4$ & $<1.54$ \\
			$\gamma_{a}$ & unconstrained $(1.376)$ & $>1.26$ \\
			$\gamma_{s}$ & unconstrained $(-2.252)$ & $-3.03(-3.76)_{-2}^{+0.57}$ \\
			$F_{z,1}$ & $0.409(0.371)_{-0.048}^{+0.023}$ & $0.379(0.400)_{-0.032}^{+0.023}$ \\
			$F_{z,2}$ & $>0.28$ & $0.317(0.275)_{-0.052}^{+0.025}$ \\
			$F_{z,3}$ & $>0.18$ & $0.190(0.164)_{-0.036}^{+0.02}$ \\
			$F_{z,4}$ & $0.091(0.105)_{-0.045}^{+0.027}$ & $0.075(0.065)_{-0.032}^{+0.022}$ \\
			$F_{\rm UV}$ & $>0.42$ & unconstrained $(0.02532)$ \\
			\hline
		\end{tabular}}
	\caption{Mean (best-fit) and $\pm 1\sigma$ errors of the cosmological and astrophysical parameters reconstructed in the models considered in this work ($\Lambda$CDM and iDM$\nu$) from the analysis of the \textit{baseline}+\lya dataset combination. Upper bounds are given at 95\% CL.}
	\label{tab:full}
\end{table}

\begin{table}
	\def\arraystretch{1.2}
	\scalebox{0.95}{
		\begin{tabular}{|l|c|c|c|c|}
			\hline
			& \multicolumn{2}{|c|}{$\Lambda$CDM} & \multicolumn{2}{|c|}{iDM$\nu$} \\
			\hline
			& \makecell{\textit{baseline} \\ } & \makecell{\textit{baseline}+ \\ \lya} & \makecell{\textit{baseline} \\ } & \makecell{\textit{baseline}+ \\ \lya} \\
			\hline
			Planck TTTEEE & $2770.30$ & $2771.61$ & $2770.00$ & $2778.81$ \\			
			Planck lensing & $8.84$ & $8.76$ & $8.85$ & $9.20$ \\
			BAO & $5.76$  & $5.21$ & $5.59$  & $5.56$ \\
			\lya &  $-$ & $39.55$ & $-$  & $32.20$ \\			
			\hline
			total $\chi^2$  & $2784.90$ & $2825.13$ & $2784.43$ & $2825.77$ \\
			\hline
	\end{tabular}}
	\caption{Best-fit $\chi^2$ per experiment (and total) for the models considered in this work ($\Lambda$CDM and iDM$\nu$) when fit to different dataset combinations.}
	\label{tab:chi2}
\end{table}

\begin{figure*}[t]
	\centering
	\includegraphics[width=0.9\linewidth]{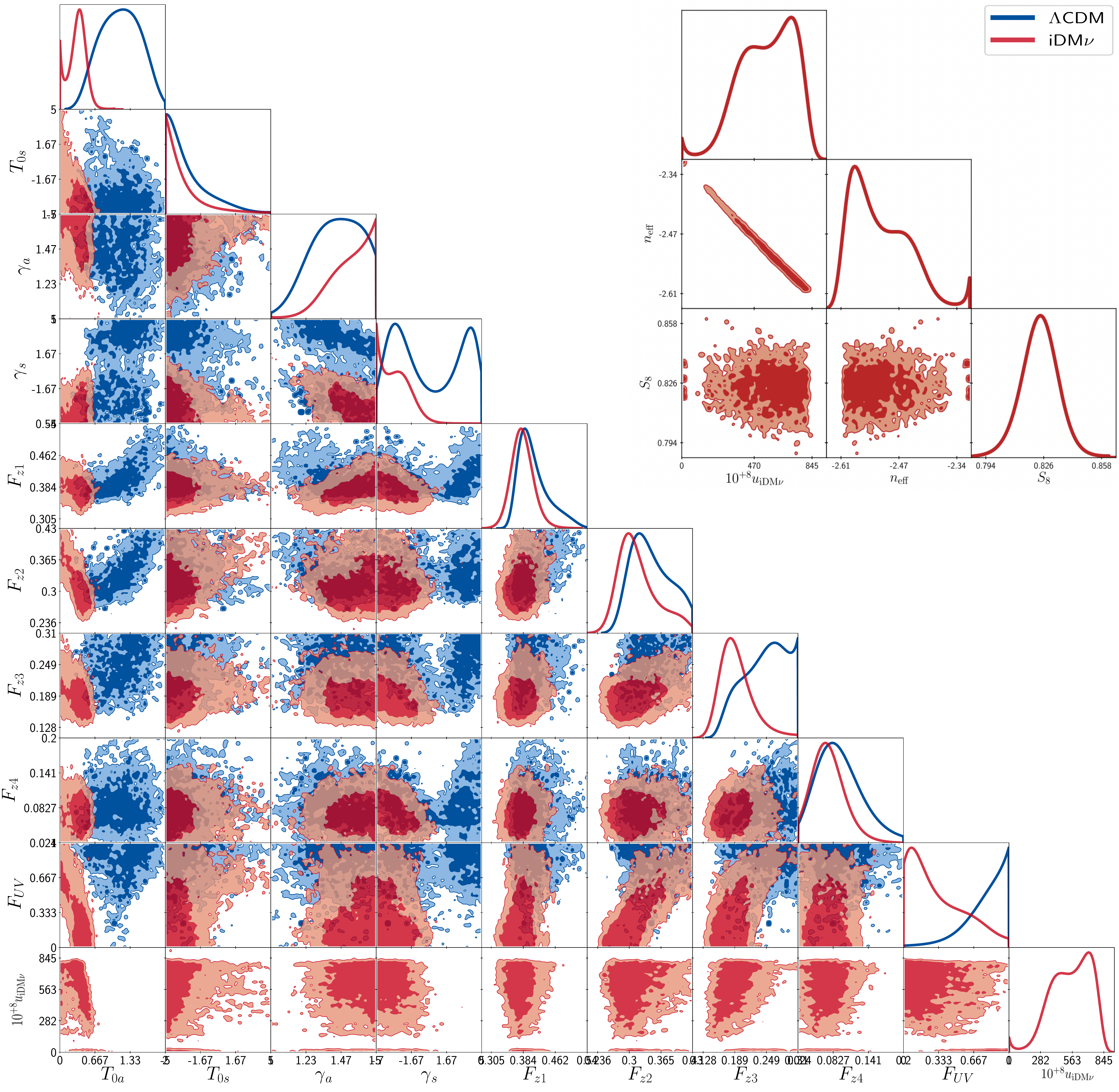}
	\caption{1D and 2D posterior distributions (68\% and 95\% CL) of the astrophysical (nuisance) and cosmological parameters for the different models considered in this work ($\Lambda$CDM and iDM$\nu$) assuming the \textit{baseline}+\lya dataset combination. In the production of the plot, the number of bins has been intentionally increased to avoid any artificial smoothing of the posteriors.}
	\label{fig: MCMC_res_nuisance}
\end{figure*}

Given their very non-Gaussian behaviour, the 1D and 2D posterior distributions of the astrophysical parameters (and $u_{\rm iDM\nu}$) are also shown in Fig. \ref{fig: MCMC_res_nuisance} to facilitate the comparison between the $\Lambda$CDM and iDM$\nu$ models. We note that, to avoid biasing the interpretation of the results, we have used a higher binning than normal. In the aforementioned figure it is possible to observe that all nuisance parameters are in rather good agreement between the two models ($\Lambda$CDM and iDM$\nu$), with the exception of $T_{0,a}$ which is lower in the iDM$\nu$ model (with a $1.8\sigma$ discrepancy). Additionally, no strong degeneracy can be seen between $u_{\rm iDM\nu}$ and any of the nuisance parameters. Moreover, we note that some of the parameters shown in the figure display contours whose bounds seem to be prior-driven, both in the $\Lambda$CDM and iDM$\nu$ scenarios. This appears to be the case also in \cite{Murgia:2017lwo} for some of these quantities, and assessing the extent to which these prior choices (for which we closely follow \cite{Murgia:2017lwo, Archidiacono2019Constraining}) affect the results is left for future work.

In the same figure (top right sub-panel), we also show the posterior distributions of the $\{u_{\rm iDM\nu}, n_{\rm eff}, S_8\}$ planes to highlight on the one hand the strong degeneracy between $u_{\rm iDM\nu}$ and $n_{\rm eff}$, and on the other hand the absence thereof between $u_{\rm iDM\nu}$ and $S_8$ (see main text for an explanation on why this is the case). We do not show the corresponding $\Lambda$CDM predictions there as they are either uninformative or already displayed in Fig. \ref{fig: MCMC_res}.

\newpage
\bibliography{bibliography}{}

\end{document}